\documentclass[a4paper,11pt]{article}
\usepackage{pos}

\usepackage[utf8]{inputenc}
\usepackage{amsmath,amscd}
\usepackage{listings}
\usepackage{caption}
\usepackage{dsfont}
\usepackage{slashed}
\usepackage{color}
\usepackage[caption=false]{subfig}

\usepackage{paralist}
\usepackage{placeins}

\usepackage{enumitem}
\usepackage{bm}
\usepackage{xcolor,colortbl}
\usepackage{multirow}

\usepackage{subfig}
\usepackage{color,graphicx,slashed,hyperref}
\usepackage[utf8]{inputenc}

\title{$B$ anomalies and muon $g-2$ from Dark Matter}

\author[a]{Giorgio Arcadi}
\affiliation[a]{Dipartimento di Scienze Matematiche e Informatiche, Scienze Fisiche e Scienze della Terra, Universita degli Studi di Messina, Via Ferdinando Stagno d'Alcontres 31, I-98166 Messina, Italy}
\author*[b]{Lorenzo Calibbi}
\affiliation[b]{School of Physics, Nankai University, Tianjin 300071, China}
\author[c]{Marco Fedele}
\affiliation[c]{Institut f\"ur Theoretische Teilchenphysik, Karlsruhe Institute of Technology, D-76131 Karlsruhe, Germany}
\author[d]{Federico Mescia}
\affiliation[d]{Dept.~de F\'{\i}sica Qu\`antica i Astrof\'{\i}sica, Institut de Ci\`encies del Cosmos (ICCUB), Universitat de Barcelona, Mart\'i i Franqu\`es 1, E-08028 Barcelona, Spain}

\emailAdd{giorgio.arcadi@unime.it}
\emailAdd{calibbi@nankai.edu.cn}
\emailAdd{marco.fedele@kit.edu}
\emailAdd{mescia@ub.edu}

\abstract{Motivated by the result of the Muon g-2 experiment and the long-standing anomalies in semileptonic $B$ meson decays, we systematically build a class of minimal models that can address both experimental results thanks to the contributions of a set of new fields that include a thermal Dark Matter candidate. This talk is mainly based on Refs.~\cite{Arcadi:2021glq,Arcadi:2021cwg}.}

\FullConference{%
  *** The European Physical Society Conference on High Energy Physics (EPS-HEP2021), ***\\
  *** 26-30 July 2021 ***\\
  *** Online conference, jointly organized by Universität Hamburg and the research center DESY ***
}

\begin{document}
\maketitle

\section{Introduction and motivations}

The FNAL Muon g-2 experiment~\cite{Abi:2021gix} confirmed the long-standing discrepancy between theoretical prediction~\cite{Aoyama:2020ynm} of the anomalous magnetic moment of the muon, $(g-2)_\mu$, and the previous measurement performed at BNL~\cite{Bennett:2006fi}. The combination of the two experiments deviates from the SM prediction by $4.2\sigma$. 
The compatibility of the two measurements convincingly excludes the possibility that the discrepancy is due to a statistical fluctuation or 
some overlooked systematical effects in the old BNL experiment~\cite{Bennett:2006fi}. 
The only possible explanations are (i)~an underestimation of the leading hadronic contribution (from hadronic vacuum polarisation) using the data-driven dispersive approach, as the recent BWM lattice result may suggest~\cite{Borsanyi:2020mff}; (ii) the presence of additional new physics (NP) contributions.

The persistent and coherent pattern of anomalies reported in semileptonic $B$ meson decays of the kind $b\to s \mu\mu$ also seems to point to a NP sector coupling preferably to muons. In particular, LHCb has recently released an updated measurement of the theoretically
clean lepton flavour universality (LFU) ratio
$R_K = \text{BR}(B\to K\mu^+\mu^-)/\text{BR}(B\to K e^+ e^-)$ reporting a $3.1\sigma$ discrepancy with the SM prediction~\cite{Aaij:2021vac}.

Assuming that they are hints of NP, both the muon $g-2$ discrepancy and the $B$ anomalies require new fields coupling to muons at scales $\lesssim \mathcal{O}(100)$~TeV~\cite{Allwicher:2021jkr,DiLuzio:2017chi}. 
Therefore, we find it  natural to seek a common explanation of the two phenomena.
 Since Dark Matter (DM) is probably the most compelling call for NP, we are interested in extensions of the Standard Model (SM) that can also account for it. Our goal is to systematically build the simplest models that can, simultaneously, (i)~address the $B$ anomalies, (ii)~explain the muon $g-2$ discrepancy, (iii)~provide a candidate of thermal-relic DM.
We aim at models that are minimal in terms of the number of new fields and their properties\,---\,quantum numbers, number of sizeable couplings, etc.\,---\,but also in the sense that the NP contributions to semileptonic $B$ decays and to the muon $g-2$ are ``induced'' by DM. In other words, the DM field is not added \emph{ad hoc} but directly enters the relevant Feynman diagrams. 

In order to fulfill the above assumptions in combination with the requirement of DM stability, our NP fields should not mix with SM fields, a condition that can be enforced by a discrete or continuous symmetry. Hence NP contributions to both muon $g-2$ and $b\to s\mu\mu$ arise at one loop (with only NP fields running in the loops) as in the general framework considered in~\cite{Gripaios:2015gra,Arnan:2016cpy,Arnan:2019uhr}.

\begin{figure}[b]
\centering
\includegraphics[width=0.90\textwidth]{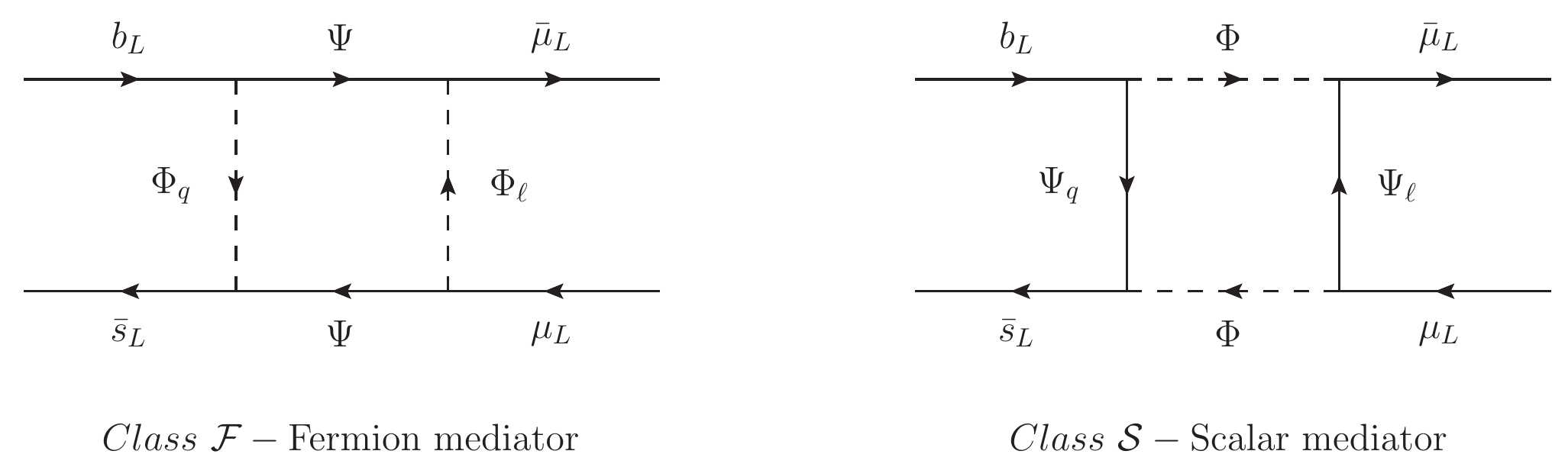}
\caption{Basic diagrams providing a contribution to $b\to s\mu\mu$
involving only left-handed SM fields. 
}
\label{fig:boxes}
\end{figure}

\section{Systematic approach to DM and $B$ anomalies}

Global fits to the $b\to s\mu\mu$ data indicate that a satisfactory fit of the $B$ anomalies is possible in presence of NP coupling to left-handed (LH) fermions only, see e.g.~\cite{Geng:2021nhg,Hurth:2021nsi,Alguero:2021anc,Ciuchini:2021smi}. 
Thus, an elegant and minimal setup to account for the $B$ anomalies consists into extending the SM spectrum with three new fields\,---\,either two scalars and one fermion or the other way round\,---\,whose quantum numbers under the SM gauge group allow for couplings with LH leptons ($L_i$) and LH quarks ($Q_i$), as shown in Figure~\ref{fig:boxes}. This class of models is described by either of the following Lagrangians:
\begin{align}
 \mathcal{L}_\mathcal{F}~ \supset ~ \Gamma^Q_i\,\bar Q_i\, P_R \Psi \, \Phi_q  + \Gamma^L_i \,\bar L_i\, P_R \Psi \, \Phi_\ell  
\,, ~\quad~
\mathcal{L}_\mathcal{S} ~\supset ~ \Gamma^Q_i\, \bar Q_i \,P_R \Psi_q \, \Phi + \Gamma^L_i\, \bar L_i\, P_R \Psi_\ell \, \Phi  \,,
\end{align}
where $\Psi_x$ are new vectorlike fermions, $\Phi_x$ are new scalars, and the labels $\mathcal{F},\,\mathcal{S}$ indicate whether the field coupled with both quarks and leptons\,---\,that we dub ``flavour mediator''\,---\,is a fermion $\Psi$ or a scalar $\Phi$. In order to highlight the minimal ingredients required by the $B$ anomalies, we only consider couplings to muons, and second- and third-generation quarks.
Besides SM gauge symmetries, the above Lagrangians are also invariant under a $Z_2$\,---\,or a global $U(1)$\,---\,symmetry in order to ensure DM stability provided it is the lightest NP particle.

\begin{figure}[t]
    \centering
    \includegraphics[width=0.42\linewidth]{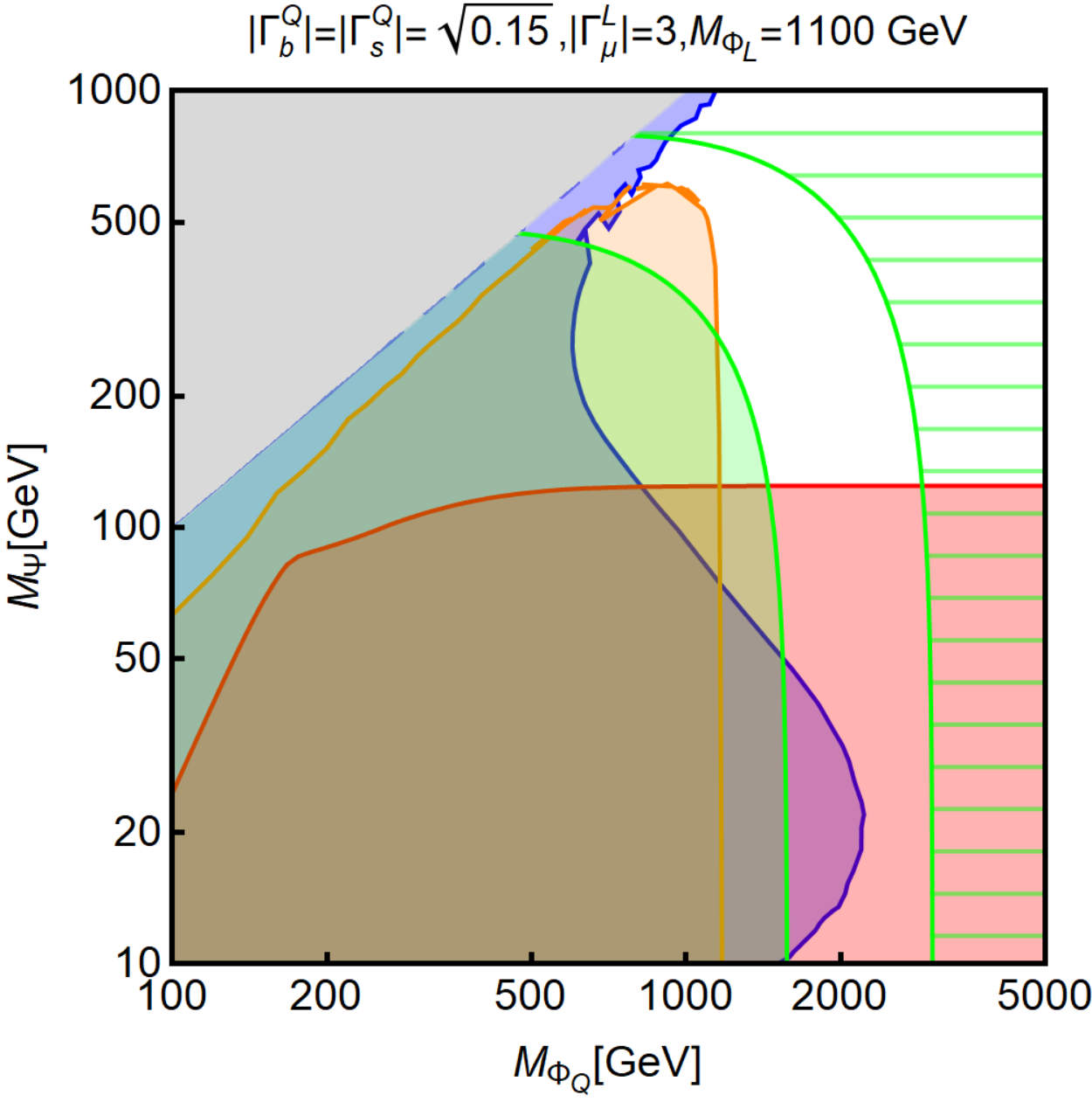}
    \caption{{Summary of constraints for the model dubbed $\mathcal{F}_{IA}$ in Ref.~\cite{Arcadi:2021glq} with DM being the Majorana fermion $\Psi$. The coloured regions are excluded by DM relic density (red), direct detection (blue), LHC searches (orange) and $B$ physics (green). The green hatched region corresponds to a SM-like contribution to $b\to s\mu\mu$, while in the white area the $B$ anomalies can be explained at the $2\sigma$ level. See Ref.~\cite{Arcadi:2021glq} for details.}}
    \label{fig:DMB}
\end{figure}

Requiring that, in order to provide a viable DM candidate, at least one of the new fields features a colourless and electrically neutral component (with null hypercharge if fermionic), a limited number of gauge quantum numbers assignments for the three new fields, and hence of possible models, is found. These were listed in Ref.~\cite{Arcadi:2021glq}, where we performed a systematic study, considering  both spin alternatives, and both cases of real/complex (Dirac/Majorana) scalar (fermion) DM.
A viable fit of the $B$ anomalies not in conflict with bounds from DM phenomenology is obtained if the following conditions are satisfied.
\begin{itemize}
    \item DM has to belong to a field directly interacting with muons with a sizeable coupling, $|\Gamma_\mu^L|\gtrsim 2$. This ensures an efficient DM annihilation into muons through t-channel diagrams, while a good fit of the $B$ anomalies can be achieved for a moderate value of the couplings to quarks, $\Gamma_s^Q\Gamma^Q_b$, thus evading stringent constraints from $B_s$ mixing. 
    \item The DM should be an $SU(2)_L$ singlet. Otherwise, DM annihilations into gauge boson pairs would be so efficient that the correct relic density would be achieved for multi-TeV DM masses, thus outside the region where a viable fit of $B$ anomalies can be achieved. Furthermore, stringent bounds from searches for disappearing tracks at LHC exist, expecially in the case DM belongs to a fermion triplet.
    \item The DM particle must be a Majorana fermion or a real scalar. If this were not the case, a viable fit of the $B$ anomalies would be ruled out by direct detection searches for DM, due to a very large photon penguin contribution to the DM-nucleon scattering cross section. 
\end{itemize}

As an illustration of these general conclusions, we show in Figure~\ref{fig:DMB} the combined constraints on a model with Majorana DM, featuring the following $SU(3)_c\times SU(2)_L\times U(1)_Y$ quantum numbers: $\Phi_q~({\bf 3},{\bf 2},1/6)$, $\Phi_\ell~({\bf 1},{\bf 2},-1/2)$, $\Psi~({\bf 1},{\bf 1},0)$. This is one of the simplest examples of a model\,---\,previously studied in Ref.~\cite{Cerdeno:2019vpd}\,---\,that successfully addresses the $B$ anomalies via interactions of a particle that can account for 100\% of the observed DM abundance.

\section{Adding the muon $g-2$}

\begin{figure}[!b]
\centering
\includegraphics[width=0.4\textwidth]{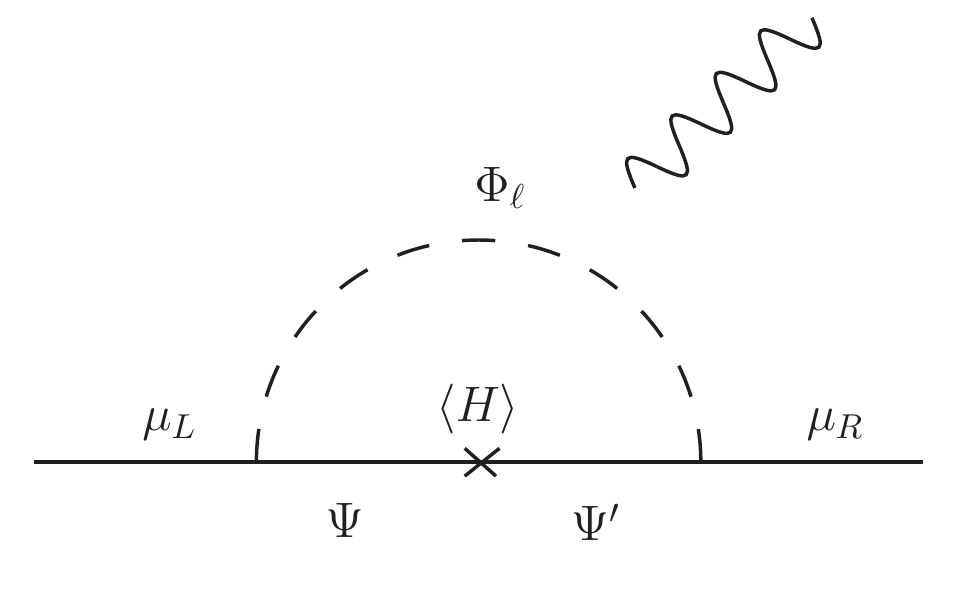}
\hspace{1cm}
\includegraphics[width=0.4\textwidth]{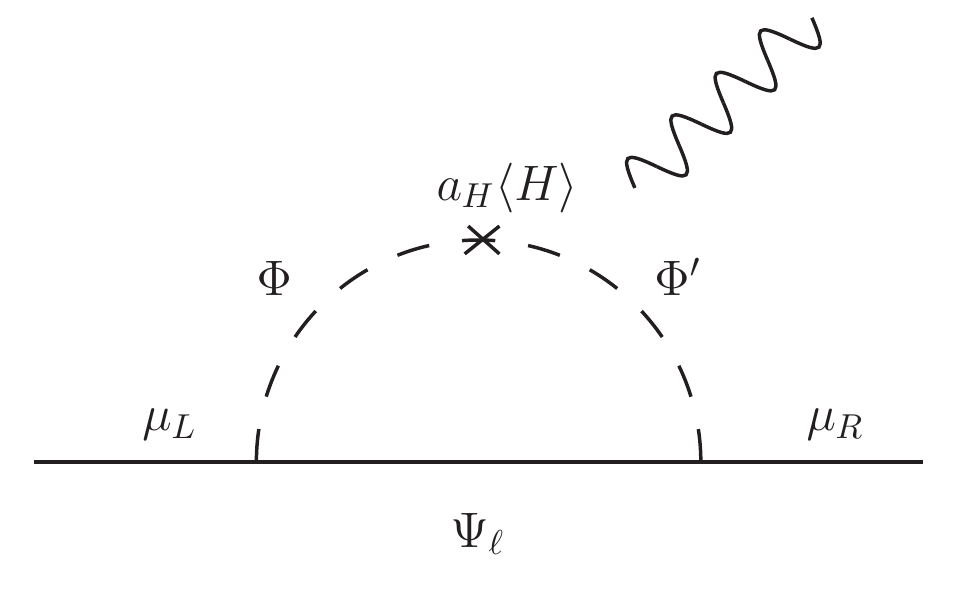}
\caption{Diagrams giving chirally-enhanced contributions to $(g-2)_\mu$.
}\label{fig:g-2}
\end{figure}

The minimal set of models illustrated in the previous section does also contribute to $(g-2)_\mu$, as penguin diagrams involving the subset of NP fields coupled to the muons induce dipole operators of the type:
\begin{equation}
    {\cal L} ~\supset~ \frac{e\, v}{8\pi^2}\, C_{\mu\mu}\left(\bar\mu_L \sigma_{\mu \nu} \mu_R\right)\, F^{\mu\nu} + {\rm h.c.} 
\quad \Rightarrow \quad a^\text{NP}_\mu = \frac{{{m_{\mu}}v}}{2 \pi^2}\, \mbox{\rm Re} (C_{\mu \mu}),
\end{equation}
where $a_\mu\equiv (g-2)_\mu/2$.
The proportionality of the operator's coefficient to the Higgs vev $v$ makes it explicit that, following from gauge invariance, a flip of the chirality of the muon\,---\,hence a Higgs insertion\,---\,is necessary. Being the NP fields coupled only with LH muons, an external mass insertion on a muon line would be the only option, causing $a^\text{NP}_\mu$ to be suppressed by the small muon Yukawa coupling,~i.e.~$C_{\mu\mu}\propto y_\mu$. While a value of $a^\text{NP}_\mu$ compatible with the experimental measurement could be still achieved, this would come at the price of low masses of the NP fields\,---\,in the 100-200~GeV range\,---\,that would not be compatible with the constraints from LHC searches and DM density outside very tuned regions of the parameter space~\cite{Calibbi:2018rzv}. To overcome this problem we need to raise the minimal number of NP fields from three to four, requiring that two of them mix through a Higgs insertion. This induces diagrams such as those shown in Figure~\ref{fig:g-2} that can lead to `chirally-enhanced' contributions to $(g-2)_\mu$, that is, not suppressed by a chirality flip $\propto y_\mu$. We then redefine our classes of models as:
\begin{align}
\text{Class }\mathcal{F}:&\quad&\text{either}\quad\{\Phi_q,\,\Phi_\ell,\,\Phi^\prime_\ell,\,\Psi\}\quad\text{or}\quad\{\Phi_q,\,\Phi_\ell,\,\Psi,\,\Psi^\prime\} \nonumber
\\
\text{Class }\mathcal{S}:&\quad&\text{either}\quad\{\Psi_q,\,\Psi_\ell,\,\Psi^\prime_\ell,\,\Phi\}\quad\text{or}\quad\{\Psi_q,\,\Psi_\ell,\,\Phi,\,\Phi^\prime\} \nonumber
\end{align}
Following again criteria of gauge invariance and DM stability, and considering the conditions listed at the end of the previous section, the analysis in Ref.~\cite{Arcadi:2021cwg} showed that the candidates for a combined explanation of $(g-2)_\mu$ and $B$ anomalies are limited to the models listed in Table~\ref{tab:models}.

\begin{table}[t!]
\centering
\renewcommand{\arraystretch}{1.2}
{\footnotesize
\begin{tabular}{ | c | c  c  c  c c|  }
\hline
Label & $\Phi_q/\Psi_q$ & $\Phi_\ell/\Psi_\ell$ & $\Psi/\Phi$ & $\Phi^\prime_\ell/\Psi^\prime_\ell$ & $\Psi^\prime/\Phi^\prime$ \\
\hline\hline
\rowcolor{cyan!40}
$\mathcal{F}_\text{Ia}/\mathcal{S}_\text{Ia}$
& $({\bf 3},{\bf 2},1/6)$ & $({\bf 1},{\bf 2},-1/2)$ & $({\bf 1},{\bf 1},0)$ & $({\bf 1},{\bf 1},-1)$ &  -- \\
\rowcolor{red!35}
$\mathcal{F}_\text{Ib}/\mathcal{S}_\text{Ib}$ & $({\bf 3},{\bf 2},1/6)$ & $({\bf 1},{\bf 2},-1/2)$ & $({\bf 1},{\bf 1},0)$ & -- &  $({\bf 1},{\bf 2 },-1/2)$ \\
\rowcolor{red!35}
$\mathcal{F}_\text{Ic}/\mathcal{S}_\text{Ic}$
& $({\bf 3},{\bf 2},7/6)$ & $({\bf 1},{\bf 2},1/2)$ & $({\bf 1},{\bf 1},-1)$ & $({\bf 1},{\bf 1},0)$ & --  \\
\hline\hline
\rowcolor{red!35}
$\mathcal{F}_\text{IIa}/\mathcal{S}_\text{IIa}$
& $({\bf 3},{\bf 1},2/3)$ & $({\bf 1},{\bf 1},0)$ & $({\bf 1},{\bf 2},-1/2)$ & $({\bf 1},{\bf 2 },-1/2)$ & -- \\
\rowcolor{cyan!40}
$\mathcal{F}_\text{IIb}/\mathcal{S}_\text{IIb}$
& $({\bf 3},{\bf 1},2/3)$ & $({\bf 1},{\bf 1},0)$ & $({\bf 1},{\bf 2},-1/2)$ & --&  $({\bf 1},{\bf 1},-1)$ \\
\rowcolor{red!35}
$\mathcal{F}_\text{IIc}/\mathcal{S}_\text{IIc}$
& $({\bf 3},{\bf 1},-1/3)$ & $({\bf 1},{\bf 1},-1)$ & $({\bf 1},{\bf 2},1/2)$ & --&  $({\bf 1},{\bf 1},0)$ \\
\hline\hline
\rowcolor{red!35}
$\mathcal{F}_\text{Va}/\mathcal{S}_\text{Va}$
& $({\bf 3},{\bf 3},2/3)$ & $({\bf 1},{\bf 1},0)$ & $({\bf 1},{\bf 2},-1/2)$ & $({\bf 1},{\bf 2},-1/2)$ & -- \\
\rowcolor{cyan!40}
$\mathcal{F}_\text{Vb}/\mathcal{S}_\text{Vb}$
& $({\bf 3},{\bf 3},2/3)$ & $({\bf 1},{\bf 1},0)$ & $({\bf 1},{\bf 2},-1/2)$ & --&  $({\bf 1},{\bf 1},-1)$ \\
\rowcolor{red!35}
$\mathcal{F}_\text{Vc}/\mathcal{S}_\text{Vc}$
& $({\bf 3},{\bf 3},-1/3)$ & $({\bf 1},{\bf 1},-1)$ & $({\bf 1},{\bf 2},1/2)$ & --&  $({\bf 1},{\bf 1},0)$ \\
\hline
\end{tabular}
}
\caption{Minimal sets of fields  fulfilling all requirements. The fields are denoted by their transformation properties under, respectively, ($SU(3)_c$, $SU(2)_L$, $U(1)_Y$). Models in cyan feature singlet DM, models in red have singlet-doublet mixed DM.
\label{tab:models}}
\end{table}

\begin{figure}[t]
    \centering
    \subfloat{\includegraphics[width=0.42\linewidth]{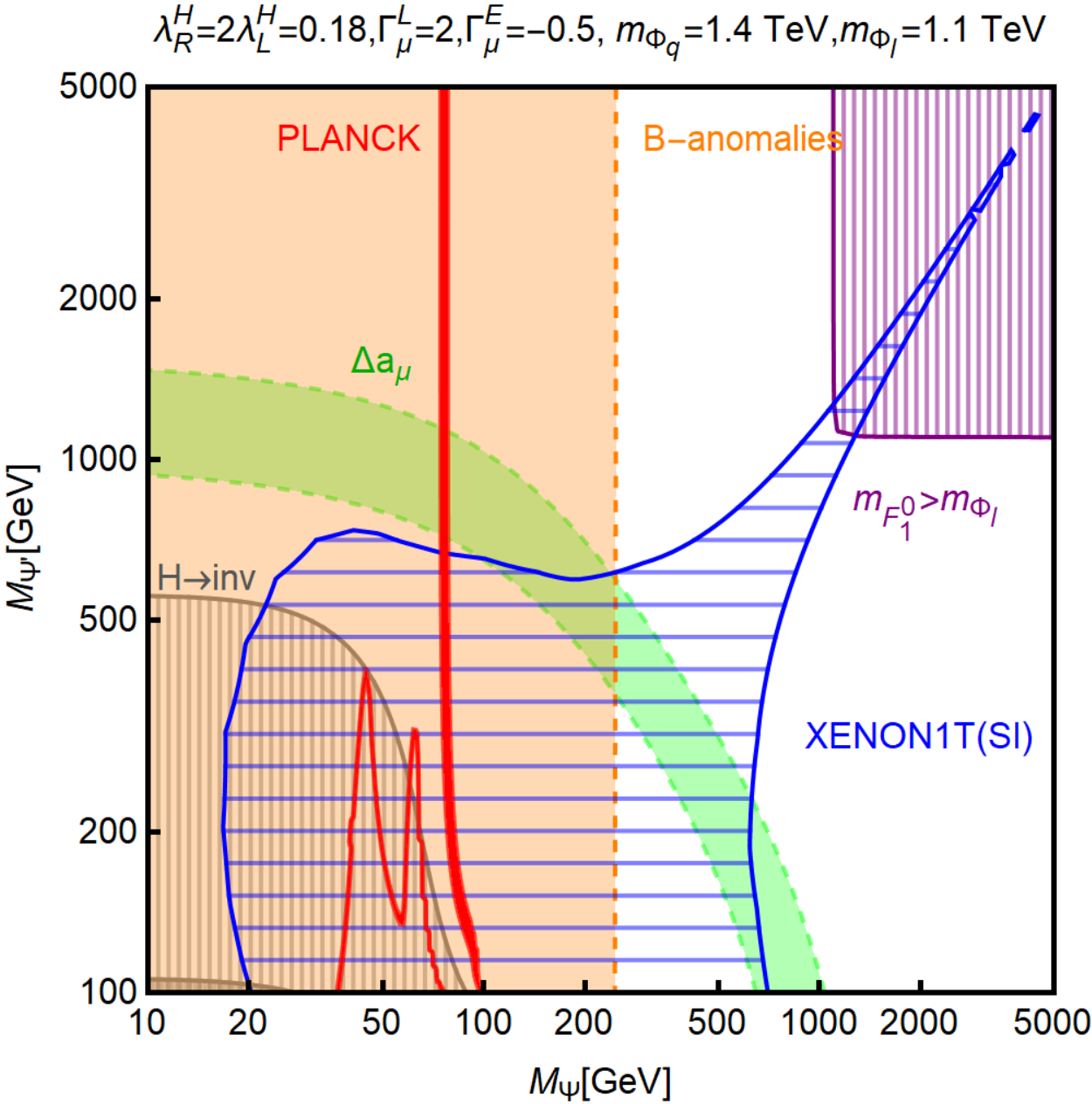}}\hspace{1cm}
    \subfloat{\includegraphics[width=0.42\linewidth]{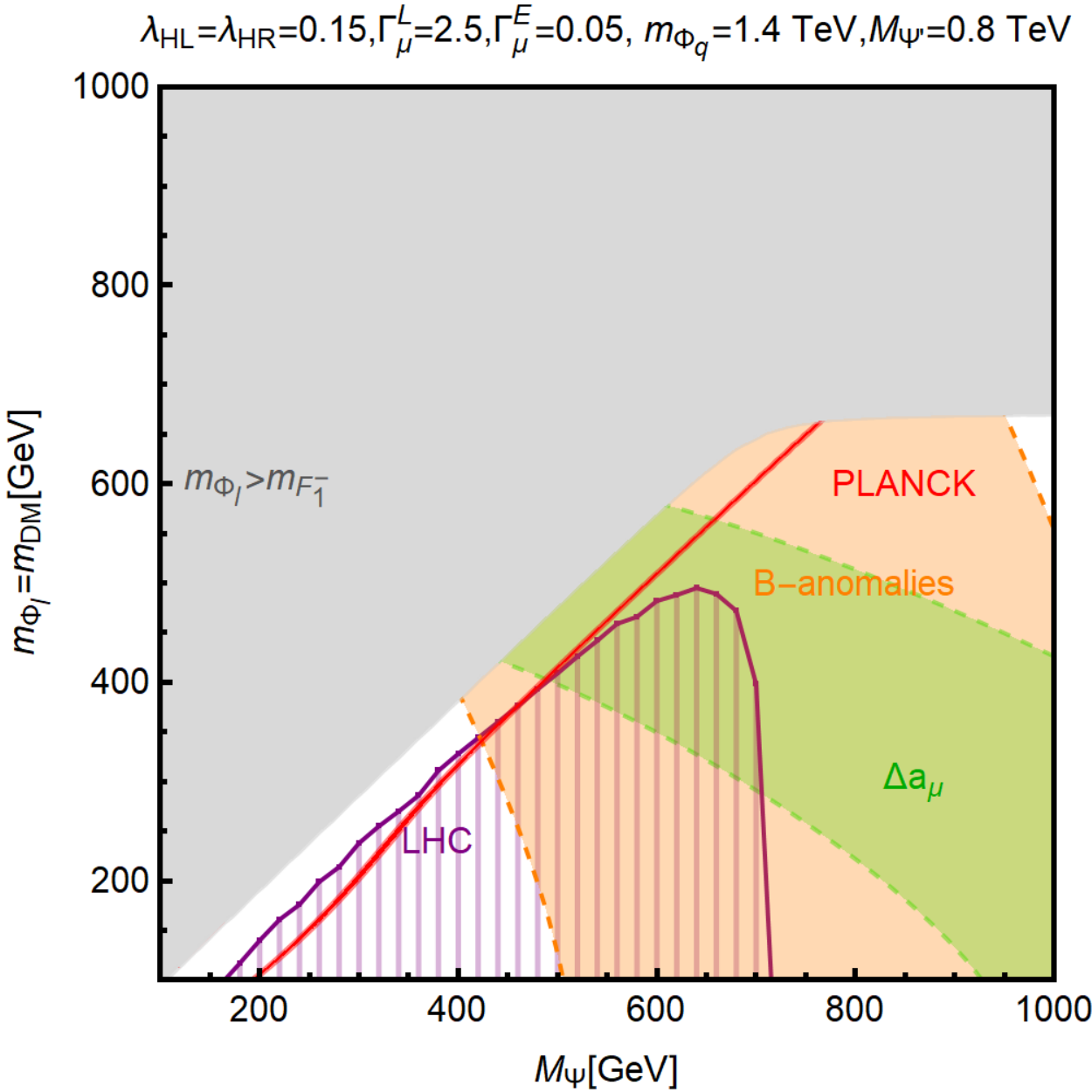}}
    \caption{Combined results for two benchmark choices of the parameters of the models $\mathcal{F_{IB}}$ (left plot) and $\mathcal{F}_{IIB}$ (right plot). The red isocontours indicate the correct DM relic density obtained from thermal freeze-out. The green and orange regions provide good fits of, respectively, the $g-2$ and $B$ anomalies. The hatched regions are excluded by direct detection (blue), the Higgs invisible width (gray) and LHC searches (purple). See~\cite{Arcadi:2021cwg} for details.}
    \label{fig:g2mod}
\end{figure}

In Figure~\ref{fig:g2mod}, we show, as illustrative examples, the two models $\mathcal{F}_{IB}$ (Majorana DM) and $\mathcal{F}_{IIB}$ (real scalar DM). 
In the first case, the chirality flip is realised by mixing of the singlet fermion $\Psi$ with a $SU(2)_L$ doublet $\Psi^{'}$. The DM candidate is the lightest neutral mass eigenstate, so that the model is an extended versions of so-called singlet-doublet DM model, see e.g.~\cite{Calibbi:2015nha}. As shown in the left panel of the figure, it is possible to achieve the correct relic density (red isocontour) and, at the same time, a viable fit of $(g-2)_\mu$ and $B$ anomalies (green and orange regions respectively). Further experimental constraints from direct detection (hatched blue region) and the invisible width of the Higgs (hatched gray region) are evaded for $M_{\Psi}/M_{\Psi^{'}}\ll 1$, corresponding to the case of singlet-like DM. In the case of the $\mathcal{F}_{IIB}$ model the DM  candidate is instead a pure singlet state (a real scalar) and the mass mixing needed for $(g-2)_\mu$ is realised by two fermionic states that mediate its annihilations. As apparent from the right panel of Figure~\ref{fig:g2mod}, also for this second model we can have a viable interpretation of the anomalies as well as the correct relic density. In this case we need to invoke sizable coannihilations between the DM and the NP fermions. As DM is a real scalar without coupling with coloured charged states, direct detection DM searches do not constrain this model. A sizable portion of the viable parameter spaces is instead excluded by LHC searches for muons and missing energy~\cite{Aad:2019vvi}, interpreted in terms of the production of the electrically-charged new fermions, followed by decays into muons and DM.

\section{Conclusions and outlook}
Following Refs.~\cite{Arcadi:2021glq,Arcadi:2021cwg}, we showed that one can systematically build minimal models simultaneously addressing the muon $g-2$ and the $B$ anomalies through loops involving a thermal DM candidate that can account for 100\% of the observed DM abundance.
This can be achieved by introducing only four new fields, at the price of a large coupling to LH muons ($\gtrsim2$) and a (moderate) chiral enhancement of the $(g-2)_\mu$ contribution.
We remark that these are not meant to be
``realistic'' models, rather the minimal ingredients that a fully-fledged theory may need to incorporate\,---\,e.g.~large muon couplings imply a Landau pole below $\approx 2500$~TeV.
Interestingly, as illustrated by the examples shown in Figures~\ref{fig:DMB} and~\ref{fig:g2mod},
these minimal solutions seem to be testable by future direct detection experiments and/or LHC searches, and, in the long run, they are definitely in the reach of a high-energy muon collider~\cite{AlAli:2021let}.
\acknowledgments 
LC would like to thank the organisers of EPS-HEP2021 and the conveners of the Dark Matter session for the opportunity of presenting our work, and acknowledges support by the National Natural Science Foundation of China under the grant No.~12035008. FM acknowledges financial support from the State Agency for Research of the Spanish Ministry of Science and Innovation through the ``Unit of Excellence Mar\'ia de Maeztu 2020-2023'' award to the Institute of Cosmos Sciences (CEX2019-000918-M) and  from PID2019-105614GB-C21 and  2017-SGR-929 grants. The work of MF is supported by the Deutsche Forschungsgemeinschaft (DFG, German Research Foundation) under grant  396021762 - TRR 257, ``Particle Physics Phenomenology after the Higgs Discovery''.


\bibliography{bibliography}
\bibliographystyle{JHEP}
          
          
\end{document}